\documentstyle[12pt]{article}

\begin{document}

\title{Calculations of the Effective Permitivity of a Periodic Array of Wires and
the Left-Handed Materials.}
\author{N. Garcia and E.V. Ponizovskaya \\
Laboratorio de F\'{\i }sica de Sistemas Peque\~{n}os y Nanotecnolog\'{\i }a,%
\\
C.S.I.C., c/Serrano 144, 28006 Madrid}
\date{\today}
\maketitle

\begin{abstract}
We present finite-differences time- domain calculations of the effective
permitivity of a periodic array of Cu strip wires that are an ingredient
needed in metamaterials for obtaining left-handed materials. We use the Cu
frequency dependent permitivity. The result is critically dependent of the
wire thickness. In particular for the wires of $0.003cm$ thickness used in
metamaterials experiments the imaginary part of the effective permitivity
dominates over the real part, which is slightly positive at $11GHz$. Wires
10 - 20 times thicker may provided a good transparent left-handed material.
\end{abstract}

Studies of photonic band gaps have shown that for a square periodic array of
wires that there is a band gap at certain wires thickness [1,3]when the
frequency of the radiation $f<c/2a$; i.e a cut-off frequency, where $c$ and $%
a$ are the speed of light and the size of the periodicity, assuming that the
medium where the wires are immersed is vacuum. This means that the effective
permitivity of the medium $\epsilon (\omega )$, $\omega =2\pi f$, is
negative. Then periodic arrays of Cu wires are used in metamaterials to
obtain negative permitivity, and split ring resonators are used to obtain
negative permeability for left-handed materials (LHM) [4,5]. In fact in a
recent paper[6] to determine the effective permitivity of metamaterials it
has been reported: '' A square array of conducting wires, based on effective
medium, is expected to exhibit the ideal frequency dependent plasmonic form
of $\epsilon =1-\omega _{co}^{2}/\omega ^{2}$, where $\omega _{co}$ is a
cut-off frequency'', the cut off given above. Clearly this argument cannot
be true in general, although we do denie that in certain cases, for
relatively thick wires it may hold, and then the $Im(\epsilon )$ may
be smaller than $Re(\epsilon )$. This is the case when the wire
thickness is much much larger that the skin depth in the wires at the given
frequency, where the calculations [1-3,6] have been done. However when the
thickness is smaller than the skin depth the medium is obviously transparent
and dielectric and the above expression for $\epsilon $ does not hold. Also
when the thickness is of the order of the skin depth there should be a
region where the expression above does not hold and in addition $Im%
(\epsilon )$ is the dominant and relevant part. There is no way to argue
against this part of physics. In fact Walser, Valanju and Valanju has stated
that for relative thin wires the losses, the imaginary part, dominates the
permitivity of the wires [7] Recently we have argued [8] that in the
experiments performed in metamaterials to obtain LHM, using wires $0.003cm$
thick we were in this last case where the imaginary part of the permitivity
dominates over the real part and then there is not LHM because of losses, no
observation of negative index of refraction can be claimed in a wedge like
sample [9]. Nevertheless Markos et al [6] have insisted that thin $0.003cm$
or thick $0.1cm$ wires should have similar permitivities, denying our
conclusions [8]. But interesting enough, all the arguments they used is that
even if they can not calculate for thin wires they expected no changes with
respect to thick wires [6,10]!!. So that nobody have calculating the
behaviour of these wires as varying the thickness. Therefore it is crucial
and very important in the problem of LHM to perform calculations of the
effective permitivity of the medium for a periodic array of Cu wires by
using the experimental $\epsilon _{Cu}(\omega )$ as the input ingredient for
the calculation in order to clarify the above problem and simulate the best
sizes for obtaining LHM.

We have performed, precisely, the above calculation using the
finite-difference time-domain method [11] and the experimental values of the
frequency dependent permitivity for Cu [12]. We show that for thin $0.003cm$
wires the imaginary part dominates and the real part is slightly positive
for $11GHz$, so that no LHM material is obtained for this frequency. However
for wires 10-20 time larger the real part of the permitivity is negative
with a small imaginary part, and then one could obtain a good transparent
LHM.

To perform the calculations we used as in the experiments [4,5] $a=0.5cm$
and the thickness $d$ of the wires was changed to see its effect on $%
\epsilon (\omega )$. The FDTD calculations Mur's first order absorbing conditions and s-polarized radiation, i.e the
electric field parallel to the wires. The $\epsilon _{Cu}(\omega )$ is
obtained by fitting the existing experimental data to obtain a plasmon
frequency $\omega _{p}$ and the damping $\gamma $, and thus we have: 
\begin{equation}
\epsilon _{Cu}(\omega )=1-\omega _{p}^{2}/(\omega ^{2}+i\gamma \omega )
\end{equation}
The inset in Fig1 shows the plot of the data [11] and the best fitting for
eq.(1) is obtained $f_{p}=8.ev$ and $\gamma =0.2$ respectively. The
imaginary and real parts are given by the dashed and continuous lines
respectively, adjusting very well to the data.

With the above permitivity for the Cu wires we have all the ingredients
needed when the wires conform a periodic array of square prisms of size $d$
in a periodic lattice of period $a=0.5cm$. Calculation for reflectivity,
transnmitivity and absorption ($R$,$T$ and $A$; $A=1-R-T$) respectively are
presented in Fig.1b to 3b for the wire thickness $d=0.1$, $0.02$ and $0.003cm
$, the later corresponding to the experimental ones [4,5] and using 3 rows
of wires. It can be seen that while the value of $R$ is practically unity
for $d=0.1cm$, this is reduced gradually for $0.02$ and $0.003cm$ and the
values of $T$ grow but specially the absorption is clearly increasing as the
thickness is reduced. This is because for $0.003$ thickness the skin depth
is $0.0005-0.001cm$, i.e. of the same order as the wire thickness and then
the absorption and losses are dominant over the real part of the permitivity.

By using the numerical simulations of Fig.1b to 3b one can fit these values
to a permitivity using an effective medium by adjusting its transmitivities
and reflectivities given by formula (4) of ref. [8] for the transmitivity,
and by similar formula for reflectivity. Least square fitting of effective
permitivity $\epsilon (\omega )$ taking the permeability equal to unity to
the numerical FDTD simulations. Results are presented in Fig. 1a to 3a and
clearly shown the following: (i) for $0.1cm$ thick wires the real part is
negative and much larger than the imaginary part of $\epsilon (\omega )$ for
all frequencies and its real part is negative; (ii) for $0.02cm$ thick wires
the real and the imaginary parts are practically equal and the real part is
negative, so that losses play a significant role; and (iii) for $0.003cm$
wires the imaginary part clearly dominates the real part that is always
positive except for the values $4GHz<f<9GHz$, in fact for $f=11GHz$ that is
the value where it is observed and band pass filter $Re(\epsilon )=0.1
$. Therefore for thin $0.003$ wires, even for $0.02cm$ thickness the
proposed existence of a cut-off plasmonic expression for the effective
permitivity is not correct. We should also say that this has been concluded
theoretically by Walser et al [7] that reach similar results simply because
for thin wires the imaginary part, losses, are dominant. The inset of Fig.3
shows the behaviour of $Re(\epsilon )$ and $Im(\epsilon )$ for
versus $d$ at the relevant frequency of $11GHz$. We also have calculated for
2 rows of wires and the results (not shown) are practically the same for the
effective permitivity.

In conclusion we have shown that the values of the effective permitivity of
a square array of wires are critically dependant of wire thickness and that
the plasmonic behaviour assumption is in general wrong but it is valid for
the relatively large thickness described here. This would suggest that by
using wires of thickness 10 to 20 times larger than used in the experiments
of metamaterials so far[4,5] it may be possible to obtain good transparent
LHM.

\pagebreak 

REFERENCES.

1. D.R. Smith, S. Schutlz, N. Kroll, M. Sigalas, K. M. Ho and C. M.
Soukoulis, Appl. Phys. Lett. 65, 645 (1994)

2. M. M. Sigalas, C.T. Chan, K. M. Ho and C.M. Soukoulis, Pjhys. Rev. B52,
11744 (1995).

3. N. Garcia, E. V. Ponizovskaya and J. Q. Xiao, Appl. Phys. Lett. 80, 1120
(2002).

4. R. A.Shelby, , D. R. Smith, S. C. Nemat-Nasser and S. Schultz, Appl.
Phys. Lett. 78, 489 (2001)

5. R. A. Shelby, D. R. Smith and S. Schultz, Science 292, 77(2001).

6. D. R. Smith, S. Schultz, P. Markos and C. M. Soukoulis, Phys. Rev. B65,
195103 (2002), P. Markos, I. Rousochatzakis and C. M. Soukoulis, preprint.

7. R. M. Walser, A. P. Valanju and P.M. Valanju, Phys. Rev. Lett. 87,
119701-1 (2001).

8. N. Garcia and M. Nieto-Vesperinas, Optics Lett. 27, 885 (2002).

9. M. Sanz, et al, Wedge-shaped Absorbing samples look lethanded: The
problem of interpreting negative refraction and its solution, Submitted to
publication,

10. A. Taflove, The finite-difference time-domain method (Artech House,
Boston 1998)

11. P. B. Johnson and R. W. Christy, Phys. Rev. B 6, 4370 (1972); J. H.
Weaver, C. Krafka, D. W. Lynch and E. E. Koch, Physics Data, Optical
Properties of Metals, ( Springer, Karlsruhe, Germany 1981). .

\pagebreak 

FIGIURE CAPTIONS

Fig.1. Inset: experimental values of $\epsilon _{Cu}(\omega )$[11] (symbols)
and fitted imaginary and real parts for $f_{p}=8eV$ and $\gamma =0.2eV$. (a) 
$Im(\epsilon )$ and $Re(\epsilon )$ ($\epsilon $ is the effective
medium permitivity) continuous and dashed lines respectively ($d=0.1cm$)
Real part is negative and dominates the imaginary part that is practically
equal to zero, very small losses. (b) FDTD calculations for reflectivity,
transnmitivity and absorption ($R$, $T$ and $A$, $A=1-R-T$). From these
simulations and using the effective medium theory the values in (a) are
obtained.

Fig.2 (a) Same as in Fig.1a for $d=0.02cm$. Real and imaginary part are of
the sam order, losses are important. Real part is negative. (b) Same as in
Fig.1b.

Fig.3 (a) Same as in Fig.1a for $d=0.003cm$. Imaginary part dominates over
the real that is positive, except for the region $4-9GHz$, but is very
small. Losses are important. This is the thickness used in metamaterials so
far [4,5] and, therefore they are not LHM at the frequencies $9-20GHz$. (b)
the same as in Fig.1b. Notice that now absorption $A$ is very important.
Inset; values of the real (dashed), left scale, and imaginary (continuous),
right scale, parts of $\epsilon $ versus $d$ at $11GHz$.

\end{document}